\documentclass[pra,twocolumn,showkeys,superscriptaddress,groupedaddress]{revtex4}

\usepackage{amsmath,amsthm,amsfonts,amssymb,times,graphicx,color}
\usepackage{multirow}
\usepackage{cases} 
\usepackage{subfigure}
\usepackage{mathtools}  
\usepackage{url}
\usepackage{hyperref} 

\usepackage{amsmath,amsfonts,amssymb,bbm}

\newcommand{\couic}[1]{}

\newcommand{\C}{\mathbb{C}}

\newcommand{\R}{\mathbb{R}}

\newcommand{\Z}{\mathbb{Z}}

\newcommand{\ii}{\mathrm i}

\newcommand{\transp}{\mathsf{T}}
\newcommand{\diag}{\textrm{diag}}

\begin{document}

\title[]{Quantum walking in curved spacetime: $(3+1)$ dimensions, and beyond}

\author{Pablo Arrighi}
\email{pablo.arrighi@univ-amu.fr}
\affiliation{Aix-Marseille Univ., CNRS, LIF, Marseille and IXXI, Lyon, France}

\author{Stefano Facchini}
\email{stefano.facchini@univ-amu.fr}
\affiliation{Aix-Marseille Univ., CNRS, LIF, Marseille, France}

\date{\today}

\begin{abstract}
A discrete-time Quantum Walk (QW) is essentially an operator driving the evolution of a single particle on the lattice, through local unitaries. Some QWs admit a continuum limit, leading to familiar PDEs (e.g. the Dirac equation).
Recently it was discovered that prior grouping and encoding allows for more general continuum limit equations (e.g. the Dirac equation in $(1+1)$ curved spacetime).
In this paper, we extend these results to arbitrary space dimension and internal degree of freedom. We recover an entire class of PDEs encompassing the massive Dirac equation in $(3+1)$ curved spacetime. This means that the metric field can be represented by a field of local unitaries over a lattice.

\end{abstract}

\maketitle

\section{Introduction}

Quantum walks (QW) \cite{BenziSucci, Bialynicki-Birula, MeyerQLGI, Kempe} are dynamics having the following characteristics: \emph{(i)} spacetime is a discrete grid; \emph{(ii)} the evolution is unitary; \emph{(iii)} the evolution is homogeneous, that is translation-invariant and time-independent, and \emph{(iv)} it is causal, meaning that information propagates at a strictly bounded speed. 

Some Quantum Computing algorithms are formulated in terms of QWs, see \cite{venegas2012quantum}. We focus here on QWs as such, as models of certain quantum physical phenomena, taking a continuum limit. Such QW-based models have a broad scope of applications:
\begin{itemize}
  \item they constitute quantum algorithms, for the efficient simulation of physical phenomena upon a quantum computer or other simulation device\cite{FeynmanQC};
  \item they constitute stable numerical schemes, even for classical computers, thereby guaranteeing convergence as soon as they are consistent \cite{arrighi2013dirac};
  \item they offer simple, discrete toy models to address questions in Foundations of physics\cite{d2014derivation,arrighi2014discrete,ArrighiKG,farrelly2014causal,farrelly2014discrete,LloydQG}.
\end{itemize}

In \cite{ArrighiCurved} we studied \emph{Paired QWs}, which on the one hand specialize general QWs described above, but in the other hand constitute a generalization of the most usual QWs found in the literature.  More specifically, \emph{(i)} the input undergoes a local, prior encoding and \emph{(ii)} the local unitary `coin' acts on larger neighborhoods. As in other QW models, the coin depends on space and time.

We showed that Paired QWs admit as continuum limit the class of PDEs of form
\begin{equation}
\partial_t \psi(t,x) = B_1 \partial_x \psi(t,x) + \frac{1}{2} \partial_x B_1 \psi(t,x) + \ii C \psi(t,x) \label{eq:ContLimitQW}
\end{equation}
with $B_1$ and $C$ hermitian and $|B_1|\leq I$. In \cite{ArrighiCurved} the spin was of dimension two (note that here and in the following we are (ab)using the word {\em spin} just as a shorthand for the internal degree of freedom, not to indicate representations of the Lorentz group). In the present work we extend this result to arbitrary, even spin dimension.
Moreover, by combining Paired QWs through operator splitting techniques, we obtain discrete models for the class of PDEs of the form
\begin{align}
\ii \partial_0 \psi &= H \psi\label{eq:ContLimitQWnD}\\
H &= \ii \sum_{i} (B^{(i)}_1 \partial_i+\frac12\sum_{i} \partial_i B^{(i)}_1) - C \nonumber
\end{align}
This class of PDEs is quite general and it includes as a special case the Hamiltonian form of the massive curved Dirac equation in $(3+1)$-dimensions \cite{de1962representations} for any bounded metric in any coordinate system, together with an electromagnetic field. Given the PDE we wish to simulate, we are able to retro-engineer the corresponding Paired QW. 

Finally we present a slightly more {\em ad-hoc} scheme that would simplify the implementation. We also relate it to Quantum Lattice Gas Automata.

\begin{figure}[t]
\centering
\includegraphics[width=8.5cm,height=4cm]{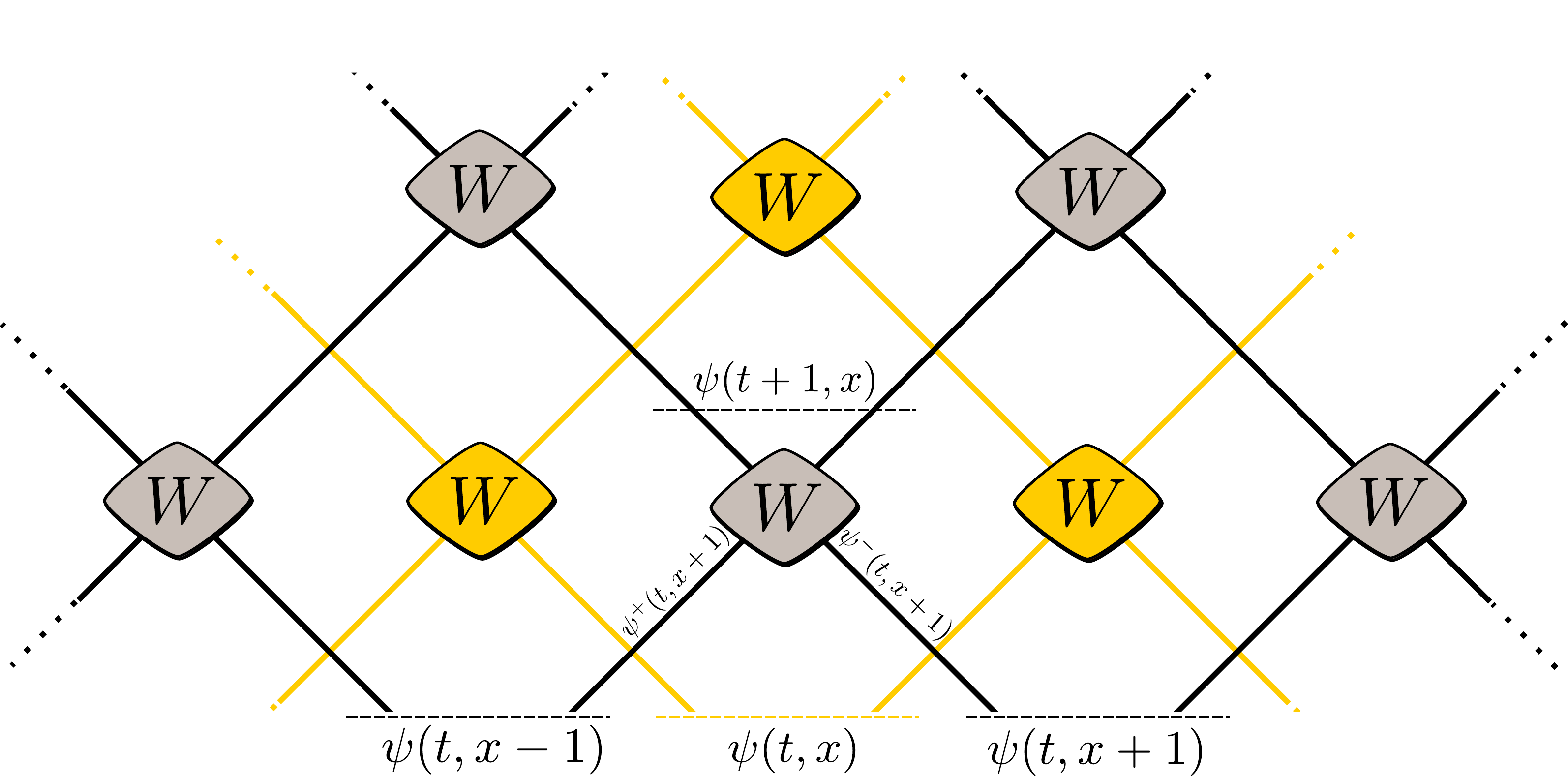} 
\caption{Usual QWs. Times goes upwards. Each site contains a $2d$-dimensional vector $\psi=\psi^+\oplus\psi^-$. Each wire propagates the $d$-dimensional vector $\psi^\pm$. These interact via the $2d\times 2d$ unitary $W$. The circuit repeats infinitely across space and time. Notice that there are two light-like lattices evolving independently.}
\label{fig:LLLat1}
\end{figure}

The results deepen the connection between QWs and the Dirac equation, first explored in \cite{BenziSucci,Bialynicki-Birula,MeyerQLGI, bracken2007free}, and further developed in \cite{dariano2012diracca,bisio2013dirac,shikano2013discrete, arrighi2013dirac, farrelly2014discrete, strauch2006relativistic}. Extension to curved spacetime was initiated in \cite{di2013quantum, di2014quantum,succi2015qwalk}.

We proceed by first extending the 1D Paired QW model to allow for arbitrary spin dimension, in Section \ref{sec:Model}. Next in Section \ref{sec:ContLimit} we compute the conditions for the continuum limit to exist, and provide solutions to these constraints. Then we extend to higher spatial dimensions, through operator splitting in Section \ref{sec:HigherSpace}. In Section \ref{sec:DiracMatching} we do the matching with the $3+1$ Curved Dirac equation. Finally, in Section \ref{sec:variations} we discuss a variation of our model, where the dimension of the local unitaries is down to the spin dimension. We put it in the form of a Quantum Lattice Gas Automaton. Some perspectives and related works are in Section \ref{sec:Discussion}.

\section{1D Paired Quantum Walks}\label{sec:Model}

\begin{figure}[t!]
\centering
\includegraphics[scale=0.25]{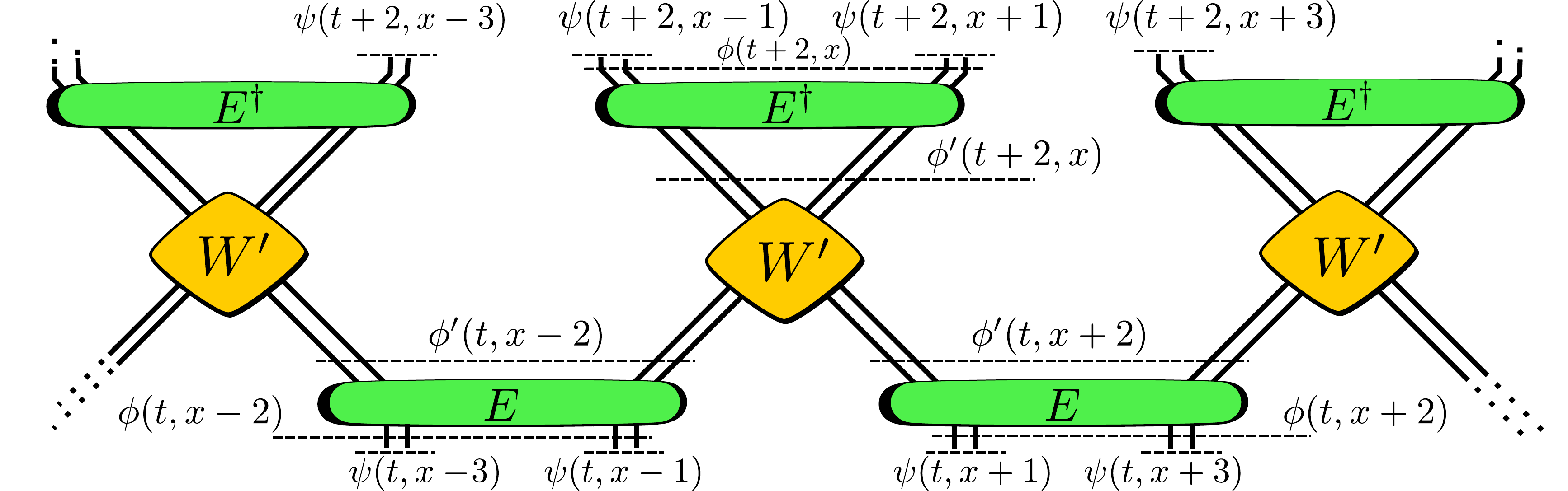} 
\caption{The input to a Paired QW is allowed to be encoded via a unitary $E$, and eventually decoded with $E^\dagger$.}
\label{fig:LLLat2}
\end{figure}

\begin{figure}[t!]
\centering
\includegraphics[width=8.5cm]{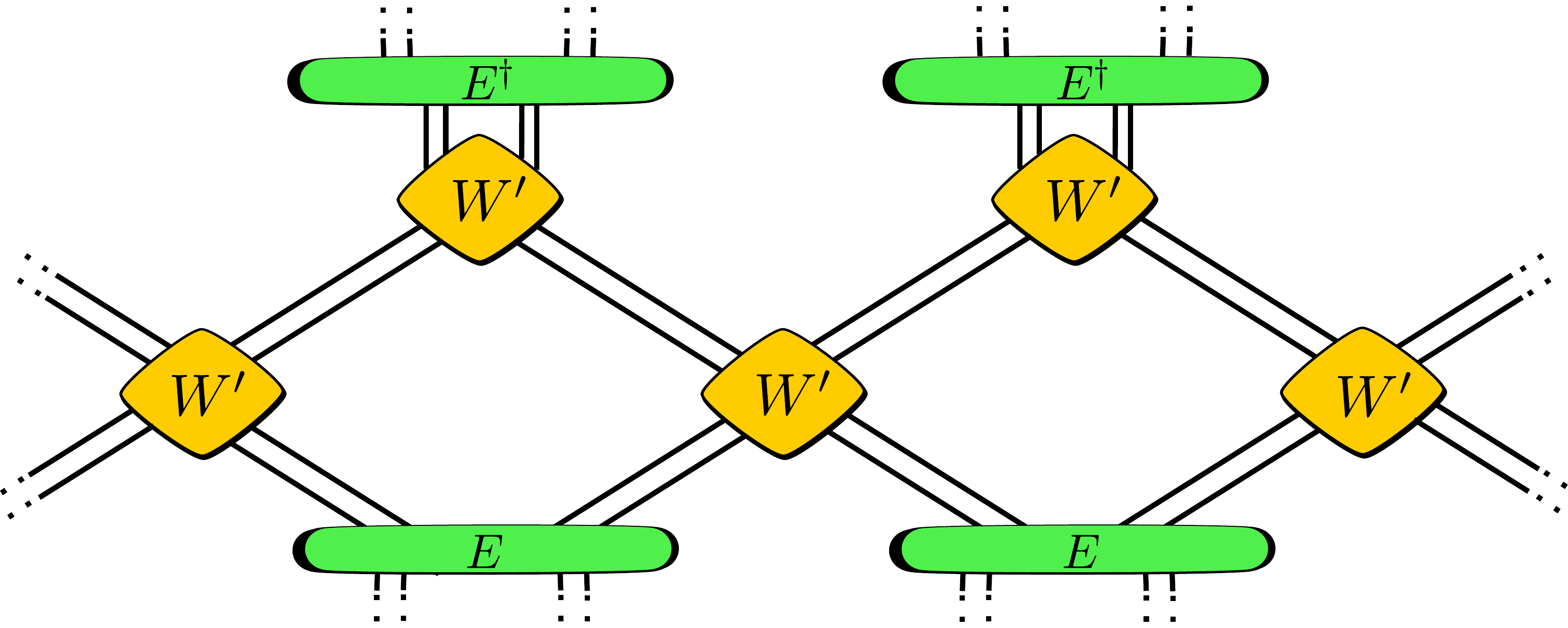} 
\caption{When the scheme is iterated, the decoding of the previous time-step cancels out with the encoding of the next time step. Thus the only relevant encoding/decoding are those of the initial input and final output. A Paired QW is therefore really just a QW, with a particular choice of initial conditions.}
\label{fig:LLLat3}
\end{figure}

Usual 1D QWs are over the space $\ell^2(\Z; \C^{s}\oplus \C^{s})$. We write $\psi(t)$ for a function taking a lattice position $x$ into the $\C^{2s}$-vector $\psi^+(t,x)\oplus\psi^-(t,x)$, with each $\psi^\pm(t,x)$ a $\C^{s}$-vector.

These QWs are obtained through the repeated application of a local unitary $W$ from $\C^{2s}$ to $\C^{2s}$, referred to as the `coin'. Hence $c=2s$ is the coin dimension or internal degree of freedom of the walker. The reason why $c$ splits as $s+s$ is because each $W(t,x)$ takes the $s$ upper components of $\psi(t,x-1)$ and the $s$ lower components of $\psi(t,x+1)$, in order to produce $\psi(t+1,x)$. Therefore the inputs and outputs of the different $W(t,x)$ are non-overlapping and the single-step evolution operator of the QW writes
\begin{equation*}
U(t) := \bigoplus_{x \in \Z} W(t,x)
\end{equation*}
where $t$ indicates the possible time dependence of the local unitaries.

Therefore usual QWs evolve two independent light-like lattices, as made clear in Fig. \ref{fig:LLLat1}. On one of the light-like lattices, the evolution is given by 
\begin{equation*}
V(t) := \bigoplus_{x \in 2\Z} W(t,x)\text{~and~}V(t+1) :=  \bigoplus_{x \in 2\Z + 1} W(t+1,x).
\end{equation*}
whilst on the other lattice everything is shifted in position.

Paired QWs were introduced in \cite{ArrighiCurved} in the particular case where $s=1$. They arise as follows. Grouping every $\psi(t,x-1)$ and $\psi(t,x+1)$ site into $\phi(t,x)=\psi(t,x-1)\oplus\psi(t,x+1)$, and applying a unitary encoding $E$ to each group, we obtain $\phi'(t,x)=E\phi(t,x)$. We may now define a QW over the space $\bigoplus_{2\mathbb{Z}} (\C^{2s}\oplus\C^{2s})$ of these encoded groups $\phi'$. The local unitary $W'$ will be from $\C^{4s}$ to $\C^{4s}$, and each $W'(t,x)$ will take the $2s$ upper components of $\phi'(t,x-2)$ and the $2s$ lower components of $\phi(t,x+2)$ in order to produce $\phi'(t+2,x)$. The inputs and outputs of the different $W'(t,x)$ are again non-overlapping and they can be applied synchronously to generate the QW evolution over the full space lattice, 
\begin{equation*}
U(t) := \bigoplus_{x \in 2\Z} W'(t,x).
\end{equation*}

In the end, each $\phi'(t+2,x)$ is decoded as $\phi(t+2,x)=E^\dagger \phi'(t+2,x)$ and ungrouped as $\phi(t+2,x)=\psi(t+2,x-1)\oplus\psi(t+2,x+1)$. Notice that this Paired QW (pictured in Figs. \ref{fig:LLLat2} and \ref{fig:LLLat3}) phrased in terms of $\phi'$ and $s'=2s$ is therefore but a subcase of the usual QW definition---from a discrete point of view at least.

When taking the continuum limit, a subtle difference shows up. Indeed, the regularity of initial condition is given in terms of $\psi(t)$, which is assumed to be smooth, i.e. $\psi(t,x)\approx\psi(t,x+1)$. It follows that the grouping $\phi(t)$ will be smooth both externally, i.e. $\phi(t,x)\approx\phi(t,x+1)$, and internally, i.e. $\phi(t,x)\approx\psi(t,x)\oplus\psi(t,x)$, which is not so usual to ask for. These reinforced regularity conditions are necessary for some Paired QWs to have a limit.

It will be useful to redefine the grouping $\phi(t,x)$ up to a unitary, as 
\begin{equation} \label{eq:Groupedwavefunction}
\phi(t,x) := \begin{bmatrix}
u(t,x) \\ d(t,x) \\ u'(t,x) \\ d'(t,x)
\end{bmatrix},
\end{equation}
with

\begin{subequations}\label{eqHad}
\begin{align}
\begin{bmatrix}
u(t,x) \\ u'(t,x)
\end{bmatrix} = (H\otimes I_s) \begin{bmatrix}
\psi^+(t,x+1) \\ \psi^+(t,x-1)
\end{bmatrix} \\ \begin{bmatrix}
d(t,x) \\ d'(t,x)
\end{bmatrix} = (H\otimes I_s) \begin{bmatrix}
\psi^-(t,x+1) \\ \psi^-(t,x-1)
\end{bmatrix}
\end{align}
\end{subequations}
where $H=\frac{1}{\sqrt{2}} \begin{pmatrix}
1 & 1 \\ 1 & -1
\end{pmatrix}$ is the Hadamard matrix and $I_s$ is the $s\times s$ identity.

This convenient choice of pre-encoding is so that in the continuum limit, to first order in the discretization parameter $\varepsilon$, we have that $u \simeq \sqrt{2} \psi^+$, $d \simeq \sqrt{2} \psi^-$, $u' \simeq \varepsilon \sqrt{2} \partial_x \psi^+$ and $d' \simeq \varepsilon \sqrt{2} \partial_x \psi^-$.

Let us focus on how $\phi_{out} := \phi(t+2,x)$ gets computed, from $\phi_{in} := \phi(t,x-2) \oplus \phi(t,x+2)$. This $\C^{4s}\oplus\C^{4s}$ to $\C^{4s}$ function is given by
\begin{align}
G &= E^\dagger(t+2,x) W'(t,x) (P' \oplus P) \nonumber \\ & \qquad (E(t,x-2)\oplus E(t,x+2)), \label{eq:FormalDefGQW}
\end{align}
where the $2s\times 4s$ projectors $P$ and $P'$ pick-up the $u,d$ (non-primed subspace) and $u',d'$ (primed subspace) coordinates, respectively. Thus
\begin{equation}
\phi(t+2,x) = G (\phi(t,x-2) \oplus \phi(t,x+2)). \label{eq:FormalGQWEq}
\end{equation}

\section{1D Continuum limit} \label{sec:ContLimit}

We will now work out the continuum limit of the Paired QW model with arbitrary spin dimension. From now on, we consider that $t$ and $x$ are continuous variables. We take $\varepsilon \in \R^+$ as the discretization parameter along every coordinate. We start by expanding Eq. \eqref{eq:FormalDefGQW}.

The expansion of the input in terms of $u,u',d,d'$, to first order in $\varepsilon$, is
\begin{equation}
\phi_{in}(t,x) \simeq \begin{bmatrix} u \\ d \\ 0 \\ 0 \end{bmatrix} \oplus \begin{bmatrix} u \\ d \\ 0 \\ 0\end{bmatrix} +
\begin{bmatrix} -2u' \\ -2d' \\ u' \\ d' \end{bmatrix} \oplus \begin{bmatrix} 2u' \\ 2d' \\ u' \\ d' \end{bmatrix}. \label{eq:phiinDev}
\end{equation}
Remember that $u'$ and $d'$ are themselves proportional to $\varepsilon$, so that the last term is proportional to $\varepsilon$.

The expansion of the output in terms of $u,u',d,d'$, to first order in $\varepsilon$, is
\begin{equation}
\phi_{out}(t,x)  \simeq
  \begin{bmatrix} u \\ d \\ 0 \\ 0 \end{bmatrix} +
  \begin{bmatrix} 2\varepsilon \partial_t u \\ 2\varepsilon\partial_t d \\ u' \\ d' \end{bmatrix}. \label{eq:phioutDev}
\end{equation}

Next we expand the walk and encoding operators, assuming that the matrix elements of $W$ and $E$ are analytic functions of $(t,x)$ and $\varepsilon$.

First, let $W' := W^{(0)}e^{\ii \varepsilon \tilde{W}}$, with $W^{(0)}$ unitary and $\tilde{W}$ hermitian. This guarantees the unitarity of $W'$. It is without loss of generality, since only its expansion to first order in $\varepsilon$ contributes to the continuum limit dynamics: 
\begin{equation}
W(t,x) \simeq W^{(0)}(t,x) + \ii \varepsilon W^{(0)}(t,x)\tilde{W}(t,x). \label{eq:Wexpansion}
\end{equation}

Then, let $E := E^{(0)}e^{\ii \varepsilon \tilde{E}}$, with $E^{(0)}$ unitary and $\tilde{E}$ hermitian. To first order in $\varepsilon$,
\begin{equation}
E(t,x) \simeq E^{(0)}(t,x) + \ii\varepsilon E^{(0)}(t,x)\tilde{E}(t,x). \label{eq:Eexpansion}
\end{equation}

We will make use of the following notation: any matrix $A \in \C^{4s\times 4s}$ will be written in block form as $A = \begin{pmatrix}
A_1 & A_3 \\ A_2 & A_4
\end{pmatrix}$, where $A_j \in \C^{2s\times 2s}$, $j=1,\ldots,4$. Let $X=\sigma_x\otimes I_{2s}$, $Y=\sigma_y \otimes I_{2s}$ and $Z = \sigma_z \otimes I_{2s}$, where $(\sigma_x,\sigma_y,\sigma_z)$ are the Pauli spin matrices. 

For any $A \in \C^{4s\times 4s}$, the following identities hold:
\begin{equation}
(P' \oplus P)(A\oplus A)(v\oplus v) = X A v \qquad \forall v\in \C^{4s} \label{eq:Simplif1}
\end{equation}
and
\begin{equation}
(P' \oplus P)(A\oplus A)(-v\oplus v) =  X Z A v \qquad \forall v\in \C^{4s}. \label{eq:Simplif2}
\end{equation}

We now discuss separately the zeroth order and the first order expansion in $\varepsilon$ of Eq. \eqref{eq:FormalDefGQW}. 

\subsection{Zeroth order} 

The zeroth order of Eq. \eqref{eq:FormalDefGQW} is
\begin{align}
\begin{bmatrix} u \\ d \\ 0 \\ 0 \end{bmatrix} &= E^{(0)\dagger} W^{(0)} (P' \oplus P)(E^{(0)}\oplus E^{(0)})
\begin{bmatrix} u \\ d \\ 0 \\ 0 \end{bmatrix} \oplus \begin{bmatrix} u \\ d \\ 0 \\ 0\end{bmatrix} \nonumber \\
&=  E^{(0)\dagger} W^{(0)} X E^{(0)} \begin{bmatrix} u \\ d \\ 0 \\ 0 \end{bmatrix},  \label{eq:ZerothOrder}
\end{align}
where we used the identity \eqref{eq:Simplif1}. This splits as
\begin{align}
\begin{bmatrix}
u \\ d 
\end{bmatrix} &= \left( E^{(0)\dagger} W^{(0)} X E^{(0)} \right)_1 \begin{bmatrix}
u \\ d 
\end{bmatrix} \label{eq:condZerothOrderA} \\
\begin{bmatrix}
0 \\ 0 
\end{bmatrix} &= \left( E^{(0)\dagger} W^{(0)} X E^{(0)} \right)_2 \begin{bmatrix}
u \\ d 
\end{bmatrix} \label{eq:condZerothOrderB}
\end{align}
Since \eqref{eq:condZerothOrderA} must hold for arbitrary $u$ and $d$, the block $1$ must be the identity. Now, since the matrix in \eqref{eq:ZerothOrder} is unitary, its rows and its columns must sum to one, thus the blocks $2$ and $3$ are zero, and \eqref{eq:condZerothOrderB} is automatically satisfied.  We still have the choice of an arbitrary unitary $U \in U(2s)$ for block $4$, to complete the matrix. Hence  
\begin{equation}
  E^{(0)\dagger} W^{(0)} X E^{(0)} = I_{2s} \oplus U, \label{eq:ZerothOrderC}
\end{equation}
where the direct sum is with respect to the non-primed subspace and the primed subspace.

\subsection{First order} 

For the first order of Eq. \eqref{eq:FormalDefGQW} a long but straightforward calculation (see Appendix \ref{app:CalculationFirstOrder}) leads to:
\begin{align}
\begin{bmatrix}2 \varepsilon \partial_t u \\ 2\varepsilon \partial_t d \\ u' \\ d'\end{bmatrix} &=  (I_{2s} \oplus U) \begin{bmatrix} 0 \\ 0 \\ u' \\ d' \end{bmatrix} + (I_{2s} \oplus U) B \begin{bmatrix} 2u' \\ 2d' \\ 0 \\ 0 \end{bmatrix} \nonumber \\ &+ \varepsilon \left\{ (2N -\ii\tilde{E})(I_{2s}\oplus U) \right. \nonumber \\ &\left.+  (I_{2s} \oplus U) (\ii\tilde{E}+2M) + T   \right\} \begin{bmatrix}
    u \\ d \\ 0 \\ 0
  \end{bmatrix}. \label{eq:ContLimitFirstOrder}
\end{align}
with
\begin{subequations}
\begin{align}
B &= E^{(0)\dagger} Z E^{(0)} \label{eq:BigB}\\
N &= (\partial_t E^{(0)\dagger}) E^{(0)} \label{eq:BigN} \\
T &= \ii E^{(0)\dagger} W^{(0)} \tilde{W} X E^{(0)}\label{eq:BigT}\\
M &= E^{(0)\dagger} Z (\partial_x E^{(0)})  \label{eq:BigM}.
\end{align}
\end{subequations}
We now focus on \eqref{eq:ContLimitFirstOrder}, studying separately its projections on the primed and on the non-primed subspaces.

\subsection{Continuum limit equation} 

On the non-primed subspace, Eq. \eqref{eq:ContLimitFirstOrder} has time derivatives in the left hand side, 
\begin{equation*}
\begin{bmatrix}
2\varepsilon \partial_t u \\ 2\varepsilon \partial_t d
\end{bmatrix} = B_1 \begin{bmatrix}
2u' \\ 2d'
\end{bmatrix} + \varepsilon ( 2 N_1 + T_1 + 2M_1  ) \begin{bmatrix}
u \\ d
\end{bmatrix}.
\end{equation*}
%where the terms containing $Q$ have simplified because it is a-hermitian.
Writing this equation in terms of $\psi(t,x)=[\psi^+(t,x),\psi^-(t,x)]^\transp$, where $\psi^\pm(t,x)$ are the original the non pre-encoded coordinates, we have
\begin{equation*}
\partial_t \psi(t,x) = B_1 \partial_x \psi(t,x) + \left(N_1 + \frac{T_1}{2}+M_1 \right) \psi(t,x).
\end{equation*}

Applying Leibniz rule to Eq. \eqref{eq:BigB}, and using \eqref{eq:BigM} we have
\begin{equation}
\partial_x B = M + M^\dagger = 2\Re M. \label{eq:ReldXBM}
\end{equation}
where $\Re M := \frac{1}{2}( M+M^\dagger)$ is the hermitian part of $M$.

From \eqref{eq:BigN}, the unitarity of $E^{(0)}$ implies:
\begin{equation}
N^\dagger = -N. \label{eq:RelNSH}
\end{equation}

From \eqref{eq:BigT},
\begin{align*}
T &= \ii E^{(0) \dagger} W^{(0)} \tilde{W} X E^{(0)} \\ &= \ii E^{(0) \dagger} W^{(0)} X E^{(0)} E^{(0) \dagger} X \tilde{W} X E^{(0)}  \\ &= \ii (I_{2s}\oplus U) E^{(0) \dagger} X \tilde{W} X E^{(0)},
\end{align*}
where we used the zeroth order condition \eqref{eq:ZerothOrderC}. Inverting,
\begin{align}
\ii E^{(0) \dagger} X \tilde{W} X E^{(0)} = (I_{2s}\oplus U^\dagger)T =
\begin{pmatrix}
T_1 & T_3 \\ U^\dagger T_2 & U^\dagger T_4
\end{pmatrix}. \label{eq:tildeWrelT}
\end{align}
The left hand term is skew-hermitian, therefore
\begin{subequations}
\begin{align}
T_1^\dagger &= -T_1  \label{eq:CondT1SH}\\
T_3 &= -T_2^\dagger U \label{eq:CondT3SH} \\
T_4^\dagger U &= -U^\dagger T_4. \label{eq:CondT4SH}
\end{align}
\end{subequations}

By splitting $M_1$ into its hermitian and skew-hermitian parts, and using equations \eqref{eq:ReldXBM}, \eqref{eq:RelNSH} and \eqref{eq:CondT1SH}, we obtain the general form of the continuum limit
\begin{equation}
\partial_t \psi(t,x) = B_1 \partial_x \psi(t,x) + \frac{1}{2} \partial_x B_1 \psi(t,x) + \ii C \psi(t,x). %\label{eq:ContLimitQW}
\end{equation}
where $C$ is an hermitian matrix given by
\begin{equation}
\ii C = N_1 + \frac{T_1}{2} + \ii\Im M_1. \label{eq:defC}
\end{equation}

\subsection{Compatibility constraints} 

On the primed subspace, Eq. \eqref{eq:ContLimitFirstOrder} becomes
\begin{align}
\begin{bmatrix}
u' \\ d'
\end{bmatrix} &= U \begin{bmatrix}
u' \\ d'
\end{bmatrix} + 2U B_2 \begin{bmatrix}
u' \\ d'
\end{bmatrix} + \varepsilon \left( 2N_2 -\ii \tilde{E}_2  \right. \nonumber \\ &\qquad + \left. \ii U\tilde{E}_2 + 2UM_2 + T_2 \right) \begin{bmatrix}
u \\ d
\end{bmatrix}. \label{eq:CompatConstraint}
\end{align}

Notice that Eq. \eqref{eq:CompatConstraint} does not contain time derivatives. These equations therefore are constraints that must be satisfied.  Indeed, recall that the continuum limit equation \eqref{eq:ContLimitQW} that we seek to obtain, is over a $\mathbb{C}^{2s}$ field, but the QW employed towards this aim is over the $\mathbb{C}^{4s}$ field obtained by grouping. Thus, as we earlier commented, the $\mathbb{C}^{4s}$ field has some internal smoothness provided by the initial regularity conditions---this must be preserved by the evolution.

In order to have nontrivial, time-dependent solutions, it must be the case that the coefficients of $[u, v]^\transp$ and $[u', v']^\transp$ vanish separately:
\begin{subnumcases}{}
U(I_{2s}+2 B_2) = I_{2s}, \label{eq:condUB} \\
2N_2  -\ii (I_{2s}-U) \tilde{E}_2 + 2UM_2 + T_2 = 0. \label{eq:condNMT}
\end{subnumcases}

\subsection{Existence of solutions}

So far we have determined the continuum limit, but only under the assumption that the constraints \eqref{eq:condUB}-\eqref{eq:condNMT} be satisfied. We now show that, given any hermitian $B_1$ and $C$, there are indeed choices of $W$ and $E$ which fulfill these constraints.

First we will show that $B_1$ along with constraint \eqref{eq:condUB} determines the zeroth order part of $E$ and $W'$.
Then, using $C$ and \eqref{eq:condNMT} we will complete the solution.

\subsubsection{Determination of $B$ and $U$}

Given $B_1$, our goal is to complete it into a $B$ of the form \eqref{eq:BigB} and satisfying \eqref{eq:condUB}. Requiring that $B$ has form \eqref{eq:BigB} is equivalent to requiring tracelessness, hermiticity, and unitarity. Expressing unitarity and hermiticity in terms of the sub-blocks gives:
\begin{align}
B_1^2 + B_2^\dagger B_2 &= I_{2s}  \label{eq:condBDiagonal} %\\ 
%B_4^2 + B_2 B_2^\dagger &= I_{2s} \label{eq:condBDiagonal2}\\
%B_2 B_1 + B_4 B_2 &= 0  \label{eq:condBAntiDiagonal} \\ 
%B_1 B_2^\dagger + B_2^\dagger B_4 &= 0   \label{eq:condBAntiDiagonal2} 
\end{align}
implying that $B_1^2<I_{2s}$. Considering then the spectral decomposition $B_1 = VDV^\dagger$, $D = \diag \{ d_1,d_2,\dots,d_{2s}\}$, the eigenvalues $d_1,d_2,\dots,d_{2s}$ must lie in $[-1,1]$ (we discuss the physical meaning of this constraint in section \ref{sec:DiracMatching}). Here is a natural solution:
\begin{align}
  B &= \begin{pmatrix} V & 0 \\ 0 & V \end{pmatrix} \overline{B} \begin{pmatrix}V^\dagger & 0 \\ 0 & V^\dagger\end{pmatrix},
  \quad \mbox{with} \label{eq:OvBV}\\ 
  \overline{B} &= \begin{pmatrix}
    D & \Lambda^\dagger \\
    \Lambda & -D
  \end{pmatrix}\label{eq:Bbar}
\end{align}
where
\begin{align*}
  \Lambda &= \diag\{-\lambda_1 e^{-\ii \eta_1}, \dots, -\lambda_{2s} e^{-\ii \eta_{2s}} \} \\
  \lambda_i &= \sqrt{1-d_i^2} \\
  \eta_i &= \arcsin |d_i|,  \qquad -\pi/2 < \eta_i < \pi/2
\end{align*}
This is indeed traceless, hermitian, and unitary. It satisfies \eqref{eq:condUB} because
\begin{align*}
  I_{2s}+2B_2&=I_{2s}+2V\Lambda V^\dagger\\
  &=V(I_{2s}+2\Lambda)V^\dagger  
\end{align*}
is unitary, using $1-2\lambda_i e^{\ii\eta_i}= -e^{-\ii 2 \eta_i}$.
The same equation gives
\begin{align*}
U&=V\diag\{-e^{\ii 2 \eta_1},\ldots,-e^{\ii 2 \eta_{2s}}\}V^\dagger.
\end{align*}

\subsubsection{Determination of $E^{(0)}$ and $W^{(0)}$}

Eq. \eqref{eq:BigB} states that $E^{\dagger (0)}$ diagonalizes $B$. Then, we can choose the columns of $E^{\dagger (0)}$ to be any complete set of normalized eigenvectors of $B$. Actually, because of the degeneracy of order $2s$ for each eigenvalue $+1$, $-1$, we could also take $\begin{pmatrix} R & 0 \\ 0 & S \end{pmatrix} E^{(0)}$ for arbitrary $R, S \in U(2s)$.

When in the special case of Eq. \eqref{eq:OvBV}, the following is an explicit solution for ${E}^{(0)}$, 
\begin{align}
  E^{(0)} &= \begin{pmatrix} V & 0 \\ 0 & V \end{pmatrix} \overline{E}^{(0)} \begin{pmatrix}V^\dagger & 0 \\ 0 & V^\dagger\end{pmatrix} \label{eq:E0}
\end{align}  
\begin{equation}
  \overline{E}^{(0)} = \frac{1}{\sqrt{2}}\begin{pmatrix}
    \overline{E}^{(0)}_1 & \overline{E}^{(0)}_3 \\
    \overline{E}^{(0)}_2 & \overline{E}^{(0)}_4
  \end{pmatrix}\label{eq:EBar0}
\end{equation}
where
\begin{align*}
  \overline{E}^{(0)}_1 &= \diag\{\nu_1^+, \dots, \nu_{2s}^+ \} \\
  \overline{E}^{(0)}_2 &= \diag\{\nu_1^-, \dots, \nu_{2s}^- \} \\
  \overline{E}^{(0)}_3 &= \diag\{-\nu_1^- e^{\ii \eta_1}, \dots, -\nu_{2s}^- e^{\ii \eta_{2s}} \} \\
  \overline{E}^{(0)}_4 &= \diag\{\nu_1^+ e^{\ii \eta_1}, \dots, \nu_{2s}^+ e^{\ii \eta_{2s}} \} \\
\end{align*}
where $\nu_i^\pm = \sqrt{1\pm d_i}$. Notice that $\overline{E}^{(0)}$ is a direct sum of $2s$ $U(2)$-unitaries, each of the form
\begin{equation}\label{eq:Edecomp}
  F_i = \begin{pmatrix} \nu_i^+ & -\nu_i^- e^{\ii \eta_i} \\
\nu_i^- & \nu_i^+ e^{\ii \eta_i} \end{pmatrix}, \quad i=1,\dots,2s
\end{equation}

Now that $E^{(0)}$ is known, $W^{(0)}$ can be computed from \eqref{eq:ZerothOrderC}. Notice that it can be written as
\begin{equation}
W^{(0)} = \begin{pmatrix} V & 0 \\ 0 & V \end{pmatrix} \overline{W}^{(0)} \begin{pmatrix}V^\dagger & 0 \\ 0 & V^\dagger\end{pmatrix} 
\end{equation}
where $\overline{W}^{(0)}$ decomposes as direct sum of $2s$ $U(2)$-unitaries, each of the form
\begin{equation}\label{eq:Wdecomp}
  W_i = F_i \begin{pmatrix} 1 & 0 \\ 0 & e^{-2\ii \eta_i} \end{pmatrix} F_i^\dagger \sigma_x, \quad i=1,\dots,2s
\end{equation}

\subsubsection{Determination of $\tilde{E}$ and $\tilde{W}$}

Having $E^{(0)}$, also determines $N_1$ and $M_1$ via Eqs. \eqref{eq:BigN} and \eqref{eq:BigM}. Finally, $T_1$ is fixed by Eq. \eqref{eq:defC} given a choice of $C$.

Notice that $\tilde{E}$ does not appear in the continuum limit, so that without loss of generality we can take it to be zero. Then, $T_2$ is also fixed by the constraint \eqref{eq:condNMT}.

The rest of $T$ can be completed by taking $T_4=0$, and $T_3$ from \eqref{eq:CondT3SH}. Finally, from \eqref{eq:tildeWrelT} we get to $\tilde{W}$,
\begin{equation}
\tilde{W} = -\ii X E^{(0)} (I_{2s}\oplus U^\dagger) T E^{(0) \dagger} X. \label{eq:DefWtilde}
\end{equation}

\subsection{Recap} 

The continuum limit of the model is given by Eq. \eqref{eq:ContLimitQW}. Given a pair of hermitian matrices $B_1$ and $C$, possibly spacetime dependent, that we wish to simulate, we are able to work out the coin $W'$ and the encoding $E$, of the QW that does the job.

%The whole procedure was programmed in {\tt sagemath}, and made available in \cite{onlinesage}.

\section{Higher spatial dimensions} \label{sec:HigherSpace}

Consider an $n+1$-dimensional spacetime with coordinates $x_0$, $x_1$,\ldots $x_n$. We can apply one above described Paired QW along dimension $x_1$, then another along $x_2$, and so on\ldots Let us investigate the result of combining such steps.

We define the Paired QWs $G^{(i)}$ as follows
\begin{align}
G^{(i)} &= E^{\dagger~(i)}(x_0+2) W^{(i)} (P' \oplus P)\label{eq:FormalDefGQWXi} \\  &(E^{(i)}(x_i-2)\oplus E^{(i)}(x_i+2)) \nonumber 
\end{align}
where $W$ is short for $W(x_0,\ldots ,x_n)$, etc., i.e. we specified only those coordinates which have been shifted. Compared to \eqref{eq:FormalDefGQW}, we will assume that the Hadamard pre-encodings of \eqref{eqHad} are part of the $E^{(i)}$, i.e. $E^{(i)}=E ((H\otimes I_{s})\oplus (H\otimes I_{s}))$. Note that the pre-encoding included in $E^{(i)}$ has now to be performed along the $x_i$ coordinate.

We proved that the continuum limit of such a QW, if it exists, has the form of \eqref{eq:ContLimitQWnD}:
\begin{align*}
\ii\partial_0 \psi &= H^{(i)} \psi\\
H^{(i)}&= \ii B^{(i)}_1 \partial_i + \ii \frac{1}{2} \partial_i B^{(i)}_1  - C^{(i)}\nonumber
\end{align*}
with $B_1$ and $C$ hermitian.
The global QW operator corresponding to $G^{(i)}$ is the unitary $U^{(i)}$ on the Hilbert space $\mathcal H = \ell^2(\Z^n) \otimes \C^{2s}$, and is such that
\begin{equation*}
  U^{(i)} \approx I - \ii 2 \varepsilon H^{(i)}.
\end{equation*}
Then, consider the discrete model given by
\begin{equation*}
  \psi(x_0+2) = U \psi =  \prod_i U^{(i)} \psi
\end{equation*}
Note that we are effectively decoding and re-encoding between each sub-step, in order to have access to the finite differences in all directions.

The continuum limit gives the equation:
\begin{align*}
  \ii \partial_0 \psi &= H \psi = \sum_i H^{(i)} \psi\\
H&= \ii \sum_i (B^{(i)}_1 \partial_i + \frac{1}{2} \partial_i B^{(i)}_1)  - C\\
C&=\sum_i C^{(i)}  
\end{align*}

\begin{figure}[t]
\centering
\includegraphics[width=\columnwidth]{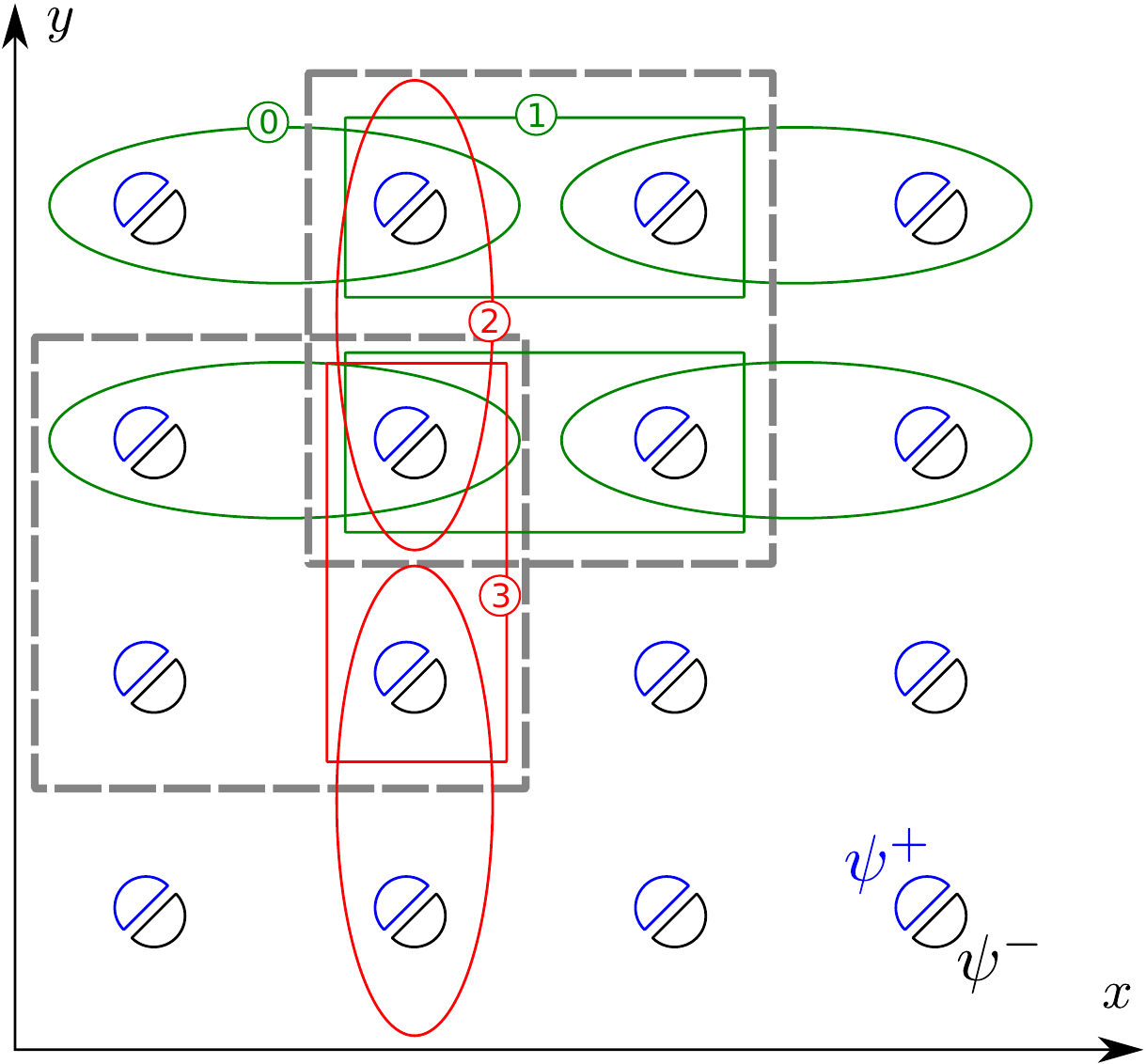}
\caption{Operator splitting two Paired QWs.}
\label{fig:opsplitting}
\end{figure}

In $(2+1)$ dimensions the model alternates one layer of $G^{(1)}$ (green in Fig. \ref{fig:opsplitting}) with one layer of $G^{(2)}$ (red in Fig. \ref{fig:opsplitting}).

\section{Recovering the $3+1$ Dirac equation} \label{sec:DiracMatching}

Consider a spacetime with metric tensor $g_{\mu\nu}$ and \emph{tetrad field} $e^{\mu}_a$, which are related to the metric as usual via $g_{\mu\nu}e^\mu_a e^\nu_b  = \eta_{ab}$, with $\eta_{ab}$ is the Minkowski metric. Assume natural units, $\hbar = c = G = 1$. In the absence of external fields, the Curved Dirac equation in Hamiltonian form \cite{de1962representations} reads 
\begin{align*}
\ii \partial_0 \psi &= H \psi\\
H &= \ii \sum_{i} (B^{(i)}_1 \partial_i+\frac12\sum_{i} \partial_i B^{(i)}_1) - C 
\end{align*}
where
\begin{align*}
B^{(i)}_1 &= - \sum_j \alpha^j \frac{e_j^i}{e_0^0} - e_0^i \\
C &= -\frac{m}{e_0^0} \beta + \frac{1}{4 e_0^0} \gamma_5 \alpha^\mu \varepsilon_{\lambda \kappa \rho \sigma} e^{\kappa \mu} e^{\rho \nu} \partial_\mu e^\sigma_\nu \\
\gamma_5 &= \ii \gamma_0\gamma_1\gamma_2\gamma_3.
\end{align*}
with $\gamma_\mu$ the usual Dirac matrices, $\beta=\gamma^0$, $\alpha^\mu = \gamma^0 \gamma^\mu$. \\
Hence, this is clearly a special case of the continuum limits of the model discussed above. These equations allow us to find the QW parameters, associated to a given metric.
The constraint that the eigenvalues of $B_1$ are $d_1, d_2, \dots, d_{2s} \in [-1,1]$ represents the finite speed of propagation on the lattice.
In practice, for any region of spacetime where the metric field is bounded, it is possible to rescale the coordinates in such a way that the physical lightcones are inside the ``causal lightcones'' of the discrete model.

\section{Lattice-gas automata form, implementation schemes}\label{sec:variations}

\begin{figure}[t]
\centering
\includegraphics[width=\columnwidth]{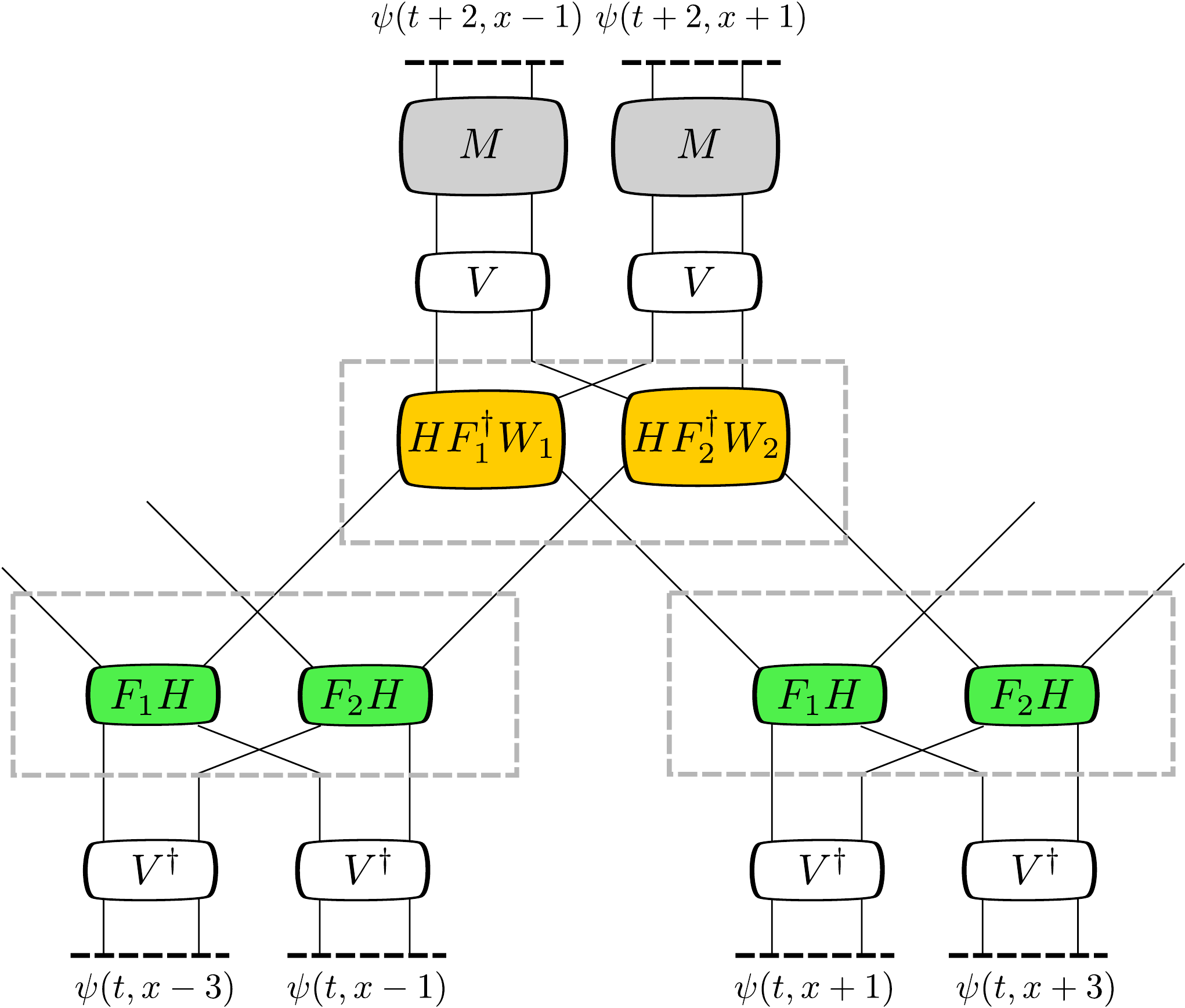}
\caption{$(1+1)$ implementation with $U(2)$ gates. Here $M = e^{2\varepsilon \ii C}$ implements the mass.}
\label{fig:2x2unitaries}
\end{figure}

\noindent {\em Lattice-gas automata form.} In $(2+1)$ dimensions the model alternates one layer of $G^{(1)}$ (green in Fig. \ref{fig:opsplitting}) with one layer of $G^{(2)}$ (red in Fig. \ref{fig:opsplitting}). At first look, this scheme does not seem to have the nice structure of Fig. \ref{fig:LLLat1} and \ref{fig:LLLat3}, namely the so called `Lattice-gas automaton', a.k.a. `partitioned cellular automaton' structure \cite{arrighi2012intrinsically}. In those structures in $(1+1)$ each cell splits into east-moving and west-moving subcells; these move (advection phase) and then undergo a local unitary (interaction phase). In $(2+1)$ this demands that each cell splits into NE/NW/SE/SW-moving subcells, which move and interact. But actually, we can gather steps $1$ and $2$ in Fig. \ref{fig:opsplitting}, and similarly steps $3$ and $0$, defining two alternating $U(8s)$ unitaries. Step $0$ acts like an initial encoding that must be decoded at the end. In this view the scheme presented in Fig. \ref{fig:opsplitting} is a natural lattice-gas-automaton-style generalization of the one in Fig. \ref{fig:LLLat3}.

\noindent {\em Implementation schemes.} Going back to the Paired QW, notice that the $E$ and $W$ of the model are apparently in $U(4s)$, i.e. twice the spin dimension. This may be a downside for two reasons. First, $U(2s)$ unitaries will be easier to implement experimentally. Second, since $2s$ is the spin dimension, a $U(2s)$-based model may seem more physical.\\
It turns out that a slight variation of the model can be constructed with only $U(2s)$ unitaries. Indeed, we have noticed that $E^{(0)}$ and $W^{(0)}$ can be decomposed as direct sum of $2s$ $U(2)$-unitaries (see Eq. \eqref{eq:Edecomp}, \eqref{eq:Wdecomp}), up to the change of basis given by the $U(2s)$ matrix $V$. Therefore in the massless case the scheme is in fact already $U(2s)$-based. An arbitrary $C$ term (see equation \ref{eq:ContLimitQW}) can be implemented with a last step of $e^{\ii \varepsilon C}$, which is again is $U(2s)$. If we proceed this way in the $(1+1)$ dimensional case (with $s=1$), and make some obvious simplification, we obtain the model given by Fig. \ref{fig:2x2unitaries}.\\ 
Up to an initial encoding, this optimized scheme decomposes into $3$ steps of local $U(2)$ coins, separated either by  partial shifts or swaps. In order to compute cell $-1$ in this scheme, $4$ cells are being looked at, with the furthest being at position $3$, hence the radius is $4$. Ignoring the independent sublattice of non-odd cells, the radius is $2$.\\
Moreover if we now operator-split this scheme to simulate the $(2+1)$ Curved Dirac equation, then up to an initial encoding, the new scheme decomposes into $6$ steps of local $U(2)$ coins, separated either by partial shifts or swaps. In order to compute cell $(-1,-1)$ in this scheme, $16$ cells are being looked at, with the furthest being at position $(3,3)$, hence the radius is $4$. Ignoring the three independent sublattices of non-odd cells, the radius is $2$.--- clearly this does not augment with spin nor space dimensions.

\section{Summary, related and future works} \label{sec:Discussion}

Paired QWs are strictly unitary, causal, local evolution operators, over a discrete spacetime---together with a prior step of encoding and a final step of decoding, see Eq. \eqref{eq:FormalGQWEq}. We have extended Paired QWs to arbirary spin dimension, showing that they admit as continuum limit all PDEs of the form \eqref{eq:ContLimitQW}. We then extended them to arbitrary space dimensions, by applying operator-splitting techniques in order to obtain all PDEs of the form \eqref{eq:ContLimitQWnD}.

This twofold extension also differentiates this work from the contributions \cite{di2013quantum, di2014quantum}\cite{succi2015qwalk}\cite{ArrighiCurved}. This finishes to prove, in all generality, that the metric field can be represented by a field of local unitaries over a lattice.

We are aware of simultaneous efforts in this direction by Debbasch et al. \cite{Debbasch2D}, who have achieved a QW for the $(2+1)$ Curved Dirac equation. Their approach is based, not on Paired QWs, but on the original stroboscopic approach of \cite{di2013quantum, di2014quantum}.  Up to an initial encoding their scheme decomposes into $8$ steps of local $U(2)$ coins, separated by partial shifts. In order to compute cell $(0,0)$ in their scheme, $25$ cells are being looked at, with the furthest being at position $(4,4)$, hence the radius is $4$. Their model is tighter, in the sense that it induces exactly the $(2+1)$ Curved Dirac equation, for synchronous coordinates $g_{00}=1$ and $g_{0i}=0$, and no more --- although it could certainly be enriched to account for electromagnetic fields. They also provide a simulation of a fermion interacting with a shear gravitational wave on a Minkowski background.

Another challenging problem is the study the underlying symmetries of the discrete model, e.g. by making explicit some form of discrete general covariance along the same lines as \cite{arrighi2014discrete}.

\begin{acknowledgments}
This work has been funded by the ANR-12-BS02-007-01 TARMAC grant, the ANR-10-JCJC-0208 CausaQ grant, and the John Templeton Foundation, grant ID 15619. The authors acknowledge helpful discussions with Marcelo Forets.
\end{acknowledgments}

\bibliographystyle{unsrt}
\bibliography{biblio}

\begin{widetext}

\appendix

\section{Calculation of the first order expansion of the discrete model} \label{app:CalculationFirstOrder}

In this section we prove Eq. \eqref{eq:ContLimitFirstOrder}, which we recall here: 
\begin{align*}
\begin{bmatrix}2 \varepsilon \partial_t u \\ 2\varepsilon \partial_t d \\ u' \\ d'\end{bmatrix} &=  (I_{2s} \oplus U) \begin{bmatrix} 0 \\ 0 \\ u' \\ d' \end{bmatrix} + (I_{2s} \oplus U) B \begin{bmatrix} 2u' \\ 2d' \\ 0 \\ 0 \end{bmatrix} + \varepsilon \left\{ (2N-\ii\tilde{E})(I_{2s}\oplus U) \right. \left.+  (I_{2s} \oplus U) (\ii\tilde{E}+2M) + T   \right\} \begin{bmatrix}
    u \\ d \\ 0 \\ 0
  \end{bmatrix}. 
\end{align*}

Recall that we want to expand 
\begin{align}
\phi_{out}(t,x) & = G~\phi_{in}(t,x), \label{eq:EvApp}
\end{align}
where
\begin{align}
G &= E^\dagger(t+2,x) W'(t,x) (P' \oplus P) (E(t,x-2)\oplus E(t,x+2)). \label{eq:EvApp2}
\end{align}
The first order expansion of the encoding and of the walk is, by definition,
\begin{align*}
E(t,x) &= E^{(0)}(t,x) + \varepsilon \ii E^{(0)}(t,x)\tilde{E}(t,x) + O(\varepsilon^2) \\
W'(t,x) &= W^{(0)}(t,x) + \varepsilon \ii W^{(0)}(t,x)\tilde{W}(t,x) + O(\varepsilon^2),
\end{align*}
hence, to first order in $\varepsilon$, the operators in \eqref{eq:EvApp2} expand to
\begin{align*}
&E^{\dagger}(t+2,x) \simeq E^{(0)\dagger} + \varepsilon \left( 2 \partial_t E^{(0)\dagger} -\ii \tilde{E}E^{(0)\dagger} \right)  \\
&W'(t,x) \simeq W^{(0)}+\varepsilon \ii W^{(0)}\tilde{W}\\
&E(t,x-2)\oplus E(t,x+2) \simeq \left( E^{(0)} - 2 \varepsilon \partial_x E^{(0)} + \ii\varepsilon E^{(0)}\tilde{E} \right) \oplus   \left( E^{(0)} + 2 \varepsilon \partial_x E^{(0)} + \ii\varepsilon E^{(0)}\tilde{E} \right) \\
\end{align*}
where in the right hand side all operators are evaluated at $(t,x)$. Recall that the first order expansions of the output and input are 
\begin{equation*}
\phi_{out}(t,x)  \simeq
  \begin{bmatrix} u \\ d \\ 0 \\ 0 \end{bmatrix} +
  \begin{bmatrix} 2\varepsilon \partial_t u \\ 2\varepsilon\partial_t d \\ u' \\ d' \end{bmatrix}, \qquad 
\phi_{in}(t,x) \simeq \begin{bmatrix} u \\ d \\ 0 \\ 0 \end{bmatrix} \oplus \begin{bmatrix} u \\ d \\ 0 \\ 0\end{bmatrix} +
\begin{bmatrix} -2u' \\ -2d' \\ u' \\ d' \end{bmatrix} \oplus \begin{bmatrix} 2u' \\ 2d' \\ u' \\ d' \end{bmatrix}.
\end{equation*}

Next we plug the previous expansions into \eqref{eq:EvApp}. Collecting all the terms of first order in $\varepsilon$, and using the identities \eqref{eq:Simplif1} and \eqref{eq:Simplif2}, we get
\begin{align*}
\begin{bmatrix} 2\varepsilon \partial_t u \\ 2\varepsilon\partial_t d \\ u' \\ d' \end{bmatrix} &=
 E^{(0) \dagger} W^{(0)} X E^{(0)} \begin{bmatrix} 0 \\ 0 \\ u' \\ d' \end{bmatrix} + E^{(0) \dagger} W^{(0)} X Z E^{(0)} \begin{bmatrix} 2u' \\ 2d' \\ 0 \\ 0 \end{bmatrix} \\ &+\left\{ \varepsilon (2\partial_t E^{(0)\dagger } -\ii \tilde{E}E^{(0) \dagger} )W^{(0)} X E^{(0)} + \ii \varepsilon E^{(0)\dagger} W^{(0)} \tilde{W} X E^{(0)} + \ii \varepsilon  E^{(0)\dagger}W^{(0)} X E^{(0)}\tilde{E} \right. \\ & +\left. 2 \varepsilon E^{(0)\dagger } W^{(0)} X Z \partial_x E^{(0)} \right\} \begin{bmatrix} u \\ d \\ 0 \\ 0 \end{bmatrix}.
\end{align*}

Next we use the zeroth order condition (cf. \eqref{eq:ZerothOrderC}), namely $ E^{(0)\dagger} W^{(0)} X E^{(0)} = I_{2s} \oplus U$, so that
\begin{align*}
\begin{bmatrix} 2\varepsilon \partial_t u \\ 2\varepsilon\partial_t d \\ u' \\ d' \end{bmatrix} &= (I_{2s} \oplus U)  \begin{bmatrix} 0 \\ 0 \\ u' \\ d' \end{bmatrix} + (I_{2s}\oplus U) \underbrace{E^{(0) \dagger} Z E^{(0)}}_{\mathrm{B }}  \begin{bmatrix} 2u' \\ 2d'\\ 0 \\ 0 \end{bmatrix} \\ &\varepsilon \left\{ \left[ 2\underbrace{(\partial_t E^{(0) \dagger})E^{(0)}}_{\mathrm{N}} -\ii\tilde{E} \right] (I_{2s}\oplus U ) + \underbrace{\ii E^{(0)\dagger} W^{(0)}\tilde{W} X E^{(0)}}_{\mathrm{T}} \right. \\ & \left. + \ii (I_{2s}\oplus U) \tilde{E} + 2 (I_{2s}\oplus U) \underbrace{E^{(0)\dagger} Z \partial_x E^{(0)}}_{\mathrm{M}} \right\}  \begin{bmatrix} u \\ d \\ 0 \\ 0 \end{bmatrix},
\end{align*}
and we get the desired result.

\end{widetext}

\end{document}